\documentclass[superscriptaddress,showpacs,amsmath,amssymb,pra]{revtex4}
\usepackage{epsfig}

\newcommand{\ket}[1]{\left| #1 \right\rangle}
\newcommand{\bra}[1]{\left\langle #1 \right|}
\newcommand\ad{{{\bf \hat{a}}_{1}^{\dag}}}
\newcommand\aponto{{{\bf \hat{a}}_{1}\cdot}}

\newcommand\pad{{{\bf \cdot\hat{a}}_{1}^{\dag}}}
\newcommand\bd{{{\bf \hat{a}}_{2}^{\dag}}}
\newcommand\bp{{{\bf \hat{a}}_{2}\cdot}}
\newcommand\bdp{{{\bf \hat{a}}_{2}^{\dag}\cdot}}

\newcommand\pbd{{{\bf \cdot\hat{a}}_{2}^{\dag}}}
\newcommand\cmat{{ m}_{1}{(t)}}
\newcommand\ccmat{{ m}^{*}_{1}{(t)}}
\newcommand\cmbt{{ m}_{2}{(t)}}
\newcommand\ccmbt{{ m}^{*}_{2}{(t)}}

\newcommand\cjat{{ j}_{1}{(t)}}
\newcommand\cjbt{{ j}_{2}{(t)}}
\newcommand\czt{{ z}{{(t)}}}
\newcommand\cczt{{ z}^{*}{{(t)}}}

\newcommand\cqt{{ q}{{(t)}}}
\newcommand\ccqt{{ q}^{*}{{(t)}}}

\newcommand{\op}[1]{{\bf \hat{#1}}}
\newcommand{\supop}[1]{{\mathcal{#1}}}
\newcommand{\de}[1]{\left( #1 \right)}
\newcommand{\choice}[2]{\stackrel{\scriptstyle{#1}}{\scriptstyle{#2}}}

\begin{document}

\title{Realistic Decoherence Free Subspaces}

\author{K.~M. \surname{Fonseca Romero}}
\affiliation{Departamento de F\'{\i}sica, Universidad Nacional,
Bogot\'a, Colombia}
\author{ S.~G. Mokarzel}
\affiliation {Departamento de F\'{\i}sica, Pontif\'{\i}cia Universidade
Cat\'olica de S\~ao Paulo, R. Marqu\^es de Paranagu\'a, 111, S\~ao Paulo, S.~P.,
Brazil} \author{M.~O. \surname{Terra Cunha}}
\affiliation{Departamento de Matem\'atica, Universidade Federal de Minas Gerais, C.P. 702, Belo Horizonte, 30123-970, Brazil}
\author{M.~C. Nemes}
\affiliation{Departamento de F\'{\i}sica--Matem\'atica, Instituto de
 F\'\i sica, Universidade de S\~ao Paulo,  C.P. 66318, 05315-970
 S\~ao Paulo, S.P., Brazil}
\affiliation{Departamento de F\'\i sica, Universidade Federal de Minas Gerais, C.P. 702, Belo Horizonte, 30123-970, Brazil}

\date{\today}
\begin{abstract}
Decoherence free subspaces (DFS) is a theoretical tool towards experimental
implementation of quantum information storage and processing. However, they
represent an experimental challenge, since conditions for their existence are
very stringent. This work explores the situation in which a system of $N$
oscillators coupled to a bath of harmonic oscillators is close to satisfy the
conditions for the existence of DFS. We show, in the Born-Markov limit and for
small deviations from separability and degeneracy conditions, that there are
{\emph{weak decoherence subspaces}} which resemble the original notion of DFS.

\end{abstract}

\pacs{03.65.Yz,03.67.Lx}

\keywords{Decoherence, Quantum Computation}

\maketitle

\section{Introduction}


The very same  mechanism responsible for the potential improvements on
computation speed using quantum mechanics, is the one which greatly hinders
immediate technical implementation. {\emph{Entanglement}} between different
subsystems is essential for the production of the states used in information
processing\cite{NC}; at the same time, as these qubits can not be completely
isolated from its environment, entanglement with the environmental degrees of
freedom is a general feature. The deleterious effect of this coupling is usually
called {\emph{decoherence}}\cite{Dec}. Therefore much effort has been devoted
to finding ways around decoherence in quantum computation, such as error
correcting codes\cite{codes}, dynamical decoupling\cite{dynadeco} and
computation in decoherence free subspaces\cite{DFS,DFS2,Zanardi}. Experimental
observations of decoherence free evolution have been
reported\cite{Science1,Science2}. Many physical implementations have been
proposed including cavity QED\cite{CQED}, ions traps\cite{Ion}, nuclear
magnetic resonance\cite{NMR} and semiconductor quantum dots\cite{QDots}. From
the theoretical point of view, recent work has been mainly on proving the
existence of DF subspaces, in general related to symmetries of the system which
are preserved by the interaction with the environment, on searching for
mechanisms of dynamical creation of DFS\cite{WL}, and on the analysis of their
robustness\cite{BLW}. More realistic models\cite{Zanardi} are scarce, and fail
to provide insight on the effects of slightly relaxing the conditions necessary
for the existence of DFS. 

In the present work we consider the case of N independent oscillators linearly
coupled to a single environment, and show that the existence of strict
decoherence free subspaces can be obtained under the following two conditions:
degeneracy of the oscillators, and separability of the coupling with the
environment. Both can be viewed as consequences of symmetries: the first
involving only the system, and the second the interaction. The exact form of the
spectral density and the temperature of the environment are immaterial in what
concerns the existence of DFS, since they really decouples from the environment.
Master equations for the evolution of the reduced density matrix of the system
are derived with and without the Born-Markov approximation, and only the
coefficients vary from one case to the other. For two independent oscillators we
solve the dynamics of the reduced density and explicit the decoherence free
subspace. Also in the case of two harmonic oscillators we study the effect of
relaxing the degeneracy and separability conditions. We verify that there is no
more DFS, but there remains a long leaved mode, which we call {\emph{weak
decoherence mode}}, and its counterbalance, a {\emph{strong decoherence mode}}
also appear. The time scales for the duration of these components are derived in
terms of the appropriate parameters. We analyze these findings in the context of
the robustness proof for DFS presented in Ref.~\cite{BLW}.

This contribution is organized as follows: in section \ref{squick} an
introduction to the concept of decoherence free subspaces is given, in section
\ref{soscillators} the model with oscillators is described, and the decoherence
free modes are exhibited. A short discussion of the classical counterpart of
decoherence free modes is made. Section \ref{smaster} is devoted to the
derivation of master equations. In section \ref{s2osc} we discuss in details the
case of two harmonic oscillators. As a simplifying tool, we introduce the notion
of {\emph{superoperator}}. The section \ref{sreal} discuss the case of small
departures from the degeneracy and separability conditions, and shows that
despite the fact that the concept of DFS is no more applicable, there remains a
{\emph{weak decoherence mode}} which can be useful for quantum information
storing. We give some concluding remarks. Some intermediate calculations have
been relegated to the appendix.

\section{A quick way to decoherence free subspaces}
\label{squick}

The notion of decoherence free subspaces (DFS) can be easily captured by
considering a special coupling between a system and its environment. Consider a
system with its autonomous Hamiltonian $\op{H}_S$, an environment described by
$\op{H}_E$, and the interaction between them given by $\op{H}_I$. The complete
Hamiltonian is then
\begin{equation}
 \op{H} = \op{H}_S + \op{H}_E + \op{H}_I.
\end{equation}
Suppose the interaction term can be written in the form $\op{H}_I
= \op{A}_S \op{B}_E$, with $\op{A}_S$ (resp.\ $\op{B}_E$) acting only on the
system (resp.\ environment) degrees of freedom. We will call this a
{\emph{separability condition}}. Any common eigenvector $\ket{a}$ of $\op{A}_S$
and $\op{H}_S$ (with eigenvalues $a$ and $h_a$) does not get entangled with the
environment, since in this case
\begin{equation}
 \begin{split}
  \op{H}\ket{a}\otimes \ket{\epsilon} &= \left( \op{H}_S \ket{a}\right) \otimes
  \ket{\epsilon} + \ket{a} \otimes \left( \op{H}_E \ket{\epsilon}\right) +
  \left( \op{A}_S \ket{a}\right) \otimes \left( \op{B}_E \ket{\epsilon}\right)
  \\
  &= \ket{a}\otimes \left( h_a + \op{H}_E + a\op{B}_E\right) \ket{\epsilon}.
 \end{split}
\end{equation}
By linearity, any common eigenspace of $\op{A}_S$ and
$\op{H}_S$ is a DFS of the system.

Degeneracy is generally originated by symmetry. Therefore, if one finds a
symmetric system where interaction with the environment preserves this symmetry,
then any eigenspace of the system is a DFS.

In the language of Ref.~\cite{DFS2}, $\op{A}_S$ is the only \emph{error
generator}, and (common) eigenspaces of (all) error generators are DFS. An
important distinction that we make is to consider as DFS only the common
eigenspaces of $\op{A}_S$ and $\op{H}_S$, \textit{i.e.:} the system Hamiltonian
should not take the state out of a DFS.

\section{A model with oscillators}
\label{soscillators}

We now present a different situation in which DFS can be achieved. The system
will consist of $N$ identical harmonic oscillators (frequency $\omega$,
annihilation operators $\op{a}_i$, we use $\hbar = 1$). The environment will be
modelled as a huge set of harmonic oscillators (frequencies $\omega _k$,
annihilation operators $\op{b}_k$). Linear coupling will be considered and the
rotating wave approximation applied. This model is both: simple enough to be
studied in details and general enough to keep the characteristic behaviour of
the problem. It is also adequated to make the link with experimental
implementations: one can consider vibronic states of $N$ ions trapped together,
a system in which (approximate) DFS has already been
demonstrated\cite{Science2,comm}, or modes of distinct cavities\cite{Switch}, or
even, for $N=2$, two degenerated modes of one cavity. The Hamiltonian to be
considered is
\begin{equation}
{\op{H}}=\omega \sum_{i=1}^N  {\op{a}_i^\dagger}{\op{a}_i} + \sum_k
\omega_k{\op{b}_k^\dagger}{\op{b}_k} + \sum_{i,k} (g_{ik}^*
{\op{a}_i^\dagger}{\op{b}_k} + g_{ik} {\op{a}_i}{\op{b}_k^\dagger}).
\label{ham}
\end{equation}
As in the previous model, we need an additional assumption on the form of the
interaction. Assume the coupling constants $g_{ik}$ can be factorized as
$G_iD_k$. This can be interpreted as supposing that all oscillators feel the
environment in the same way, possibly with only a difference in strength, which
depends only on the oscillator itself, not on the environment (the most usual
models consider $G_i = G$, which is a special form of the here proposed model).
With this factorizability hypothesis, one can rewrite the Hamiltonian
(\ref{ham}) as
\begin{equation}
{\op{H}}= \omega \sum_{i=1}^N  {\op{a}_i^\dagger}{\op{a}_i} + \sum_k
\omega_k {\op{b}_k^\dagger \op{b}_k} + \sum_{k} \left(D_{k}^* (\sum_i
G_i^* {\op{a}_i^\dagger}){\op{b}_k} + D_{k}(\sum_i G_i {\op{a}_i})  {\op{b}_k^\dagger}\right).
\end{equation}
By defining the collective operators
\begin{equation}
{\op{ A}^\dagger_1} = \frac{\sum_i G_i^*{\op{a}_i^\dagger}}{\sum_i |G_i|^2},
\quad {\op{A}_1} = \frac{\sum_i G_i{\op{a}_i}}{\sum_i |G_i|^2},
\end{equation}
it takes the form
\begin{equation}
{\op{H}}=\omega \sum_{i=1}^N  {\op{a}_i^\dagger}{\op{a}_i} + \sum_k
\omega_k{\op{b}_k^\dagger \op{b}_k} +  {\op{A}_1^\dagger}\sum_{k}
c_{k}{\op{b}_k} + {\op{A}_1}\sum_{k} c_{k}^*{\op{b}_k^\dagger},
\end{equation}
where it is clear that only the collective mode
$\op{A}_1$ is coupled to the environment (in the above formula $c_k = \sum_i
|G_i|^2 D_k^*$). One can consider the mode $\op{A}_1$ as the first of a new set
of normal modes $\left\{ \op{A}_i\right\}$, and the remaining modes thus
constitute an infinite dimensional DFS. With this new set of variables, the
Hamiltonian is finally written as
\begin{equation}
{\op{H}}=\omega \sum_{i=1}^N  {\op{A}_i^\dagger}{\op{A}_i} + \sum_k
\omega_k{\op{b}_k^\dagger \op{b}_k} +  {\op{A}_1^\dagger}\sum_{k}
c_{k}{\op{b}_k} + {\op{A}_1}\sum_{k} c_{k}^*{\op{b}_k^\dagger}.
\label{finham}
\end{equation}

While the situation in the previous section is completely quantum mechanical,
the present model  does have a classical analog, because the manipulation above
can also be done with classical oscillators. In fact, there is a very old
classical situation to which this analysis can be applied: synchronization of
pendular clocks. It is known that two clocks in the same wall tend to
synchronize in anti-phase. Each clock can be considered as an oscillator, and
their coupling to the environment can be considered in terms of the two normal
modes (phase and anti-phase modes). The in phase mode couples (much more)
strongly to the environment, and this causes the synchronization.

Another consistent analogy of the above model is with the superradiance in the
Dicke model\cite{Dicke}. In this case, $N$ two-level atoms are coupled to one
field mode. In the regime in which the atoms are collectively coupled to the
field ({\it{i.e.}}: the field does not distinguish which atom has emitted), the
radiative process can be stronger compared to individual emissions (due to
interference). The counterpart of this process is the subradiance, {\it{i.e.}}:
other collective states with weaker emission than the individual contributions
(destructive interference). DFS can thus be compared to subradiant states of
the system.

\section{Master equation}
\label{smaster}
As one usually does not have control on the environment degrees of freedom, the
natural approach to this problem is to study the reduced dynamics of the $N$
oscillators. A long but straightforward procedure\cite{alamos} can be applied to
derive the master equation 
\begin{equation}
\label{Eq:Master}
\begin{split}
\frac{d{\op{\rho}}}{dt} = &\frac{1}{i\hbar}\left[{\op{H}_0},{\op{\rho}} \right]
+\left(\lambda+\epsilon \right) \left( 2{\op{A}_1\op{\rho}\op{A}_1^\dagger}-
{\op{A}_1^\dagger \op{A}_1\op{\rho}} -{\bf \op{\rho}
\op{A}_1^\dagger\op{A}_1}\right)\\
& +\epsilon \left( 2{\op{A}_1^\dagger\op{\rho}\op{A}_1}-{\op{A}_1
\op{A}_1^\dagger\op{\rho}} -\op{\rho}{\op{A}_1 \op{A}_1^\dagger}\right),
\end{split}
\end{equation}
where
\begin{equation}
{\op{H}_0}=\hbar\omega \sum_{i=2}^N{\op{A}_i^\dagger\op{A}_i}+
\hbar(\omega+\delta){\op{A}_1^\dagger\op{A}_1}.
\end{equation}
The real functions $\lambda,\delta,\epsilon$ are implicitly defined in terms of
the auxiliary function $\eta(t)$
\begin{equation}
\label{eta}
\eta(t)=\exp\left(
-\int_0^t \lambda(t') dt'-iwt -i\int_0^t \delta(t')dt'
\right),
\end{equation}
which satisfies the integrodifferential equation
\begin{equation}
\label{Eq:Integrodif}
\dot{\eta}+i\omega \eta + \int_0^t d\tau
\sum_k |c_k|^2 {\rm e}^{i\omega_k(t-\tau)} \eta(\tau)=0,
\end{equation}
subject to the initial condition $\eta(0)=1$.
Moreover, considering the environment in thermal equilibrium, we have
\begin{equation}
\epsilon(t) =  \frac{|\eta(t)|^2}{2 }\frac{d}{dt}
\left( \sum_k
\frac {|c_k|^2 n_k (\beta)}{|\eta(t)|^2}
\left|
\int_0^t d\tau e^{-i\omega_k (t-\tau)} \eta(\tau)
\right|^2 \right),
\end{equation}
where $n_k(\beta)$ is the mean excitation number for the $k^{\text{th}}$ mode of the environment
at inverse temperature $\beta= 1/k_B T$. If the usual Born-Markov
approximations hold, then $\delta(t) =0$, $\lambda(t)= \sum_i k_i := k$,
and $\epsilon= k\bar{n}$, where $k_i$  characterize the
Markovian evolution when only the $i^{\text{th}}$ original oscillator is
coupled to the bath, and $\bar{n}$ is the environment mean number
of thermal excitations at frequency $\omega$. In this case the master equation
simplifies to
\begin{equation}
\label{Eq:BMMaster}
\begin{split}
\frac{d{\op{\rho}}}{dt} = &
-i\omega\sum_{i=1}^N \left[{\op{A}_i^\dagger\op{A}_i},{\op{\rho}}
\right] + k \left( \bar{n}+1\right) \left( 2{\op{A}_1\op{\rho}
\op{A}_1^\dagger}-{\op{A}_1^\dagger \op{A}_1\op{\rho}}
-\op{\rho}{\op{A}_1^\dagger \op{A}_1}\right)\\ & + k \bar{n}\left(
2{\op{A}_1^\dagger\op{\rho}\op{A}_1}-{\op{A}_1 \op{A}_1^\dagger\op{\rho}}
-\op{\rho}{\op{A}_1\op{A}_1^\dagger}\right).
\end{split}
\end{equation}
As one should expect by eq.~(\ref{finham}), the master equation above describes
the dissipative evolution of the collective mode $\op{A}_1$, and the independent
unitary evolution of the remaining modes $\op{A}_i$, $i\geq 2$. It also should
be noted that the damping constant $k$ of mode $\op{A}_1$ is larger (in general,
much larger, for large $N$) than the individual constants $k_i$ of the modes
$\op{a}_i$. This should be compared to the superradiance analogy discussed at
section \ref{soscillators}.

\section{Two oscillators in a dissipative environment}
\label{s2osc}

From now on, we especialize to the case of two harmonic oscillators ($N=2$).
The collective modes can thus be written as
\begin{equation}
{\op{A}_1^{\dagger}} = \cos (\theta) {\op{a}_1^{\dagger}}
                  + \sin(\theta) {\op{a}_2^{\dagger}},\quad
{\op{A}_2^{\dagger}} = -\sin (\theta) {\op{a}_1^{\dagger}}
                  + \cos(\theta) {\op{a}_2^{\dagger}},
\label{AiRai}
\end{equation}
with $\tan \theta = G_2/G_1$ (this quotient can be taken as a positive number,
if necessary, by redefining the mode $\op{a}_2$). If Markovian approximation is
made, this relation takes the form $\tan \theta = \sqrt{k_2/k_1}$. The
expression (\ref{AiRai}) can be considered as giving $\left\{
\op{A}_i^{\dagger}\right\}$  by applying a rotation operator $\supop{R}\left(
\theta \right)$ on the set $\left\{ \op{a}_i^{\dagger}\right\}$. It is usual to
call an operator which acts on operators a {\emph{superoperator}}. Thus,
$\supop{R}\left( \theta \right)$ is a {\emph{rotation superoperator}}. It is
convenient to represent superoperators using the algebraic relations among the
operators in which they act on, and a useful notation is to introduce a dot
($\bullet$) in the position where the operator to be acted on must be placed.
For example, $\left[ \op{A}, \bullet \right] \op{B} = \left[ \op{A},
\op{B}\right]$, and $\left[ \op{A}, \bullet \right]^2 \op{B} = \left[ \op{A},
\left[ \op{A}, \op{B}\right] \right]$. With this convention in mind one can
verify that
\begin{equation}
\supop{R}\left( \theta \right) =
\exp{\left\{{\theta \left[ \op{a}_1\op{a}_2^{\dagger} -
\op{a}_2\op{a}_1^{\dagger}, \bullet \right]}\right\}}.
\end{equation}

Superoperators are very useful to study time evolution. As one can define an evolution operator $\op{U}\left( t\right)$ by $\op{U}\left( t\right) \ket{\psi \left( 0\right)} = \ket{\psi \left( t\right)}$, the {\emph{evolution superoperator}} is defined by $\supop{U}\left( t\right) \op{\rho}\left( 0\right) = \op{\rho}\left( t\right)$. One completely solves the time evolution of a system by writing its evolution superoperator. The equation (\ref{Eq:Master}) can be solved by the superoperator
\begin{equation}
\label{SuperUA}
\supop{U}\left( t\right) =
e^{-iwt\left[{\op{A}_2^\dagger \op{A}_2} ,\bullet \right]}
 v e^{\left( 1-v\right) {\op{A}^\dagger_1 \bullet \op{A}_1}}
 e^{x {\op{A}^\dagger_1 \op{A}_1 \bullet}}
 e^{x^*{\bullet \op{A}^\dagger_1 \op{A}_1}}
 e^{z {\op{A}_1\bullet \op{A}^\dagger_1}}
\end{equation}
where the coefficients $v(t)$, $x(t)$ and $z(t)$ can be
given in terms of the functions $\eta(t)$, eq. (\ref{eta}), and
$\mathcal{N}(t)$,
\begin{equation}
 {\cal N}(t)=
\int_0^t d\tau \epsilon(\tau)
\left|\frac{\eta(\tau)}{\eta(t)}\right|^2,
\end{equation}
as follows
\begin{equation} \nonumber
v(t)  =  \frac{1}{1+{\cal N}(t)}, \quad
x(t)  =  \ln \frac{\eta(t)}{\sqrt{1+{\cal N}(t)}}, 
\quad z(t)  =  1-\frac{\left|\eta(t)\right|^{-2}}{1+{\cal N}(t)}.
\end{equation}
If the Markovian limit is applied, the preceding formulas reduce to
\begin{equation}
v=\frac{1}{1+\bar{n} (1-e^{-2(k_1+k_2)t})},\quad
x= \ln \frac{e^{(-i\omega-k_1-k_2) t}}{\sqrt{1+\bar{n}(1-e^{-2( k_1 +k_2)t})}},\quad
z=\frac{(\bar{n}+1)(1-e^{-2(k_1+k_2)t})}{1+\bar{n}(1-e^{-2(k_1+k_2)t)})}.
\end{equation}
The evolution superoperator $\supop{U}\left( t\right)$ can be expressed in terms of the original mode operators $\op{a}_i$ by using the rotation superoperator $\supop{R}\left( \theta \right)$ in the following way
\begin{equation} \label{SuperU}
\mathcal{U}(t) =
e^{\theta \left[{\op{a}_1 \op{a}_2^\dagger}-{\op{a}_2
\op{a}_1^\dagger},\bullet \right]}
e^{-i\omega t\left[{\op{a}_2^\dagger \op{a}_2} ,\bullet \right]}
v e^{\left( 1-v\right) {\op{a}^\dagger_1} \bullet {\op{a}_1}}
e^{x {\op{a}^\dagger_1 \op{a}_1} \bullet}
e^{x^*{\bullet \op{a}^\dagger_1 \op{a}_1}}
e^{z {\op{a}_1}\bullet {\op{a}^\dagger_1}}
e^{-\theta \left[ {\op{a}_1 \op{a}_2^\dagger}-{\op{a}_2
\op{a}_1^\dagger},\bullet \right]}.
\end{equation}

Since the second collective mode is effectively decoupled from the
environment, any density operator in the Hilbert space of this mode,
times the asymptotic density operator of the coupled collective mode,
provided it exists, will experience a unitary evolution. For simplicity
we will further on restrict ourselves to the zero temperature case.
Any density operator of the form (kets $\ket{m,n}$ refer to the original modes $\op{a}_i$):
\begin{equation}
\begin{split}
{\op{\rho}} & = \sum_{n,m} \frac{{{\rho}}_{n,m}}{n!m!}
\left( \op{A}_2^{\dagger}\right)^n
\ket{0,0}\bra{0,0}
\left( \op{A}_2\right)^m
\\
& = \sum_{n,m} \frac{{{\rho}}_{n,m}}{n!m!}
\left(-{\op{a}_1^\dagger} \sin{\theta}+{\op{a}_2^\dagger}
\cos{\theta}\right)^n \ket{0,0}\bra{0,0}
\left(-{\op{a}_1} \sin{\theta}+{\op{a}_2} \cos{\theta}\right)^m
\\ & =
\sum_{n,m,n_1,m_1} {{\rho}}_{n,m}
\frac{\sqrt{n!m!}(-\sin\theta)^{n+m-n_1-m_1}(\cos\theta)^{n_1+m_1}}
{\sqrt{(n-n_1)!n_1!(m-m_1)!m_1!}}
\\ & \qquad \qquad \quad \times
\ket{n_1,n-n_1}\bra{m_1,m-m_1},
\end{split}
\end{equation}
will be protected against dissipation and decoherence. In fact,
applying the evolution superoperator to an initial density matrix of
this form, we obtain
\begin{equation}
\begin{split}
{\op{\rho}}(t) &  =
\sum_{n,m,n_1,m_1} e^{-i\omega t (n-m)}{{\rho}}_{n,m}
\frac{\sqrt{n!m!}(-\sin\theta)^{n+m-n_1-m_1}(\cos\theta)^{n_1+m_1}}
{\sqrt{(n-n_1)!n_1!(m-m_1)!m_1!}}
\\ & \qquad \qquad\qquad\qquad \qquad \times
\ket{n_1,n-n_1}\bra{m_1,m-m_1},
\end{split}
\end{equation}
as one must expect.

Now, we use the evolution superoperator on the initial operator density
\begin{equation}\label{initialcondition}
{\op{\rho}}\left( 0\right)=\left( \cos{\alpha} \ket{1,0}+ e^{i\phi}\sin{\alpha}\ket{0,1}\right)
\left( \cos{\alpha}\bra{1,0}+e^{-i\phi}\sin{\alpha}\bra{0,1}\right),
\end{equation}
which can be viewed as a one photon Fock state of the mode given by the creation operator
\begin{equation}
\label{mode}
\op{A}^{\dagger} \left( \alpha ,\phi \right) = \cos{\alpha}\ \op{a}_1^{\dagger} +  e^{i\phi}\sin{\alpha}\ \op{a}_2^{\dagger},
\end{equation}
where $\alpha$ and $\phi$ can be compared to Stokes parameters describing
polarization.  We will obtain how the dissipative properties of the mode $\left(
\alpha, \phi \right)$ depend on these parameters. As this is a natural way for
experimentally test DFS\cite{Science2}, in the next section a more realistic
situation is discussed. The above state asymptotically approaches the rank $2$
density operator given by
\begin{equation}
\label{asympdens}
{\op{\rho}}_{t\rightarrow\infty} = P
\ket{\psi}\bra{\psi} +\left( 1-P\right)\ket{0,0}\bra{0,0}
\end{equation}
where the state $\ket{\psi}$ depends
only on the individual decay rates $k_i$, \begin{equation}
\ket{\psi} =\frac{\sqrt{k_2} \ket{1,0} -\sqrt{k_1} \ket{0,1}}{\sqrt{k_1+k_2}}
\end{equation}
and the weight $P$ of this state is given by
\begin{equation}
P = \bra{\psi}\op{\rho}\left( 0\right) \ket{\psi} =
\left|
\frac{\sqrt{k_2}\cos{\alpha}-\sqrt{k_1}e^{i\phi}\sin{\alpha}}
     {\sqrt{k_1+k_2}}
\right|^2.
\end{equation}

Observe that varying $\alpha$ and $\phi$ we can go from total preservation
to total leakage. For example, if we set $\tan(\alpha) =
\sqrt{k_1/k_2}$, and $\phi=0$ then the full state will leak to the
ground state $\ket{0,0}$, since in this case $\alpha = \theta$ and the initial photon was in the ``superradiant'' mode $\op{A}_1$. On the other hand, if we set $\tan(\alpha) =
-\sqrt{k_2/k_1}$, and $\phi=0$ then the initial state will be
exactly equal to $\ket{\psi}$ (one photon in mode $\op{A}_2$), and will persist at all
times with probability 1 (aside for an unimportant global phase). All other
combinations will go to the density operator (\ref{asympdens}), which can be
considered as an ensemble of pure state $\ket{\psi}$ with
probability $P$ and the ground state $\ket{0,0}$ with probability $1-P$. One
can define the asymptotic fidelity, $F_{\infty}\left( \alpha ,\phi \right)$,
which is the overlap between the initial and asymptotic density matrices. In the
above example it is given by
\begin{equation}
F_{\infty}\left( \alpha ,\phi
\right) = \left| \frac{(\sqrt{k_2}\cos(\alpha)-\sqrt{k_1}e^{i\phi}\sin(\alpha))
(\sqrt{k_2}\cos(\alpha)-\sqrt{k_1}e^{-i\phi}\sin(\alpha))} {k_1+k_2}\right|^2.
\end{equation}

\section{Effects of more Realistic Modeling}
\label{sreal}
We remark that the results above were obtained under a number of
assumptions, which will be relaxed below. Notice that the use
of the rotating wave approximation (RWA) is not essential in obtaining
the decoupled mode: any interaction linear in the field operators
would be as good (provided the other assumptions hold). Had we chosen
an interaction linear in the identical oscillators but nonlinear on
the environmental operators we would have obtained also a decoupled
collective mode. In these cases, however, the complication would be
only of technical nature leading to (much) more complex dynamics.

Another important hypothesis to obtain DFS is that of identical
frequencies of the original main oscillators. Of course, any
interaction between them would destroy the symmetry upon which the
existence of DFS rests. On the other hand, we have assumed that the
oscillator-environment coupling satisfies $g_{ik} = G_i D_k$, which
amounts to a separable coupling. It is not an easy task to find
realizations of such interactions in nature given its nonlocal
character. However, it might be a good approximation in special
circumstances, as \textit{e.g.} optical cavities. A particular consequence
of the separability hypothesis can be seen writing the master equation,
in the zero temperature limit,
in terms of the original oscillators (with different frequencies
for generality)
\begin{eqnarray}
\label{MasterdaSonia}
{\cal L}_0 & = &\left(-i\omega_{1} - {k_1}\right) \op{a}^{\dagger}_1\op{a}_1\bullet
+\left( i\omega_{1}-{k_1}\right) \bullet \op{a}^{\dagger}_1\op{a}_1
+2{k_1}\op{a}_1 \bullet \op{a}^{\dagger}_1 +   \nonumber\\
& & \left(-i\omega_2-{k_{2}}\right) \op{a}^{\dagger}_2\op{a}_2\bullet
+\left( i\omega_{2}-{k_{2}}\right) \bullet \op{a}^{\dagger}_2\op{a}_2
+2{k_{2}}\op{a}_2 \bullet \op{a}^{\dagger}_2 +    \nonumber\\
&&{k_{3}}\left(2\op{a}_1\bullet \op{a}^{\dagger}_2-\op{a}^{\dagger}_2\op{a}_1\bullet - \bullet \op{a}^{\dagger}_2\op{a}_1\right)
+{k_{3}}^*\left(2\op{a}_2\bullet \op{a}^{\dagger}_1-\op{a}^{\dagger}_1\op{a}_2\bullet - \bullet \op{a}^{\dagger}_1\op{a}_2\right).
\end{eqnarray}
The new quantity $k_3$ appears since we consider the same environment
interacting with both oscillators. These terms can be considered as an
interaction between the oscillators mediated by the environment. If the
separability condition is fulfilled, $\left|k_3\right| ^2 = k_1k_2$, and if the
oscillators are identical, eq.~(\ref{MasterdaSonia}) is the same as
eq.~(\ref{Eq:BMMaster}). The other limit case is to consider both oscillators
interacting independently with the environment. In this situation, the
independence of the phases of the interaction coefficients $g_{ik}$ will make
their net effect on $k_3$ null, and the eq.~(\ref{MasterdaSonia}) will just
describe two independent damped harmonic oscillators. Our interest is to study
the above equation when the conditions for the existence of DFS are almost
satisfied, {\it{i.e.}} $\left|k_3\right| ^2 \approx k_1k_2$ and $\omega _1
\approx \omega _2$. It must be noted that, by Cauchy-Schwarz
inequality,$\left|k_3\right| ^2 \leq k_1k_2$.


We should point out that we are not using the usual approach of perturbation
theory, of adding a small perturbation $\varepsilon \op{H}'$. In
Ref.~\cite{BLW}, the authors have shown that DFS are robust up to order
$\varepsilon$ in the perturbation, and all orders in time, so our approach must
be connected to second order (in $\varepsilon$) perturbation.

The explicit solution to this problem is given in the appendix. As is expected, there is no DFS without the separability and degeneracy assumptions, but if we are close to this conditions, we can obtain states much more robust to decoherence and dissipation than others. As in the previous section, consider the one photon states of eq.~(\ref{initialcondition}). Then we can define a {\emph{weak decoherence mode}} (WD), which tends to the DFS when degeneracy and separability are approximated,  and a {\emph{strong decoherence mode}} (SD) which is analogous to the superradiant mode $\op{A}_1$. We want to explore the slight deviations from separability and degeneracy, so we define $\delta k$ and $\delta \omega$ by
\begin{equation}
\delta k = \sqrt{k_1k_2} - \left| k_3\right|, \quad 2\delta \omega = \omega _2 - \omega _1,
\end{equation}
and consider $\delta \omega \ll \omega _i$ and $\delta k \ll k_i$.

As in the previous section, varying the parameters of the initial state
(\ref{initialcondition}) can be interpreted as varying the mode of the initial
photon. In the regime discussed above we obtain:
\begin{equation}
\label{wsmodes}
\op{A}^{\dagger}_{\choice{\mbox{\tiny{SD}}}{\mbox{\tiny{WD}}}} =
\frac{1}{\sqrt{k_1 + k_2}} \left( \sqrt{k_{\choice{1}{2}}} \op{a}^{\dagger}_1
\pm \sqrt{k_{\choice{2}{1}}}e^{\pm i\frac{\delta \omega}{k}}\op{a}^{\dagger}_2
\right),
\end{equation}
which must be compared to eq.~(\ref{mode}). The new
parameter $k$ is a kind of effective mean damping, and in the regime here
discussed can be considered as $k \approx \left( k_1 + k_2\right) /2$. Each mode
has its own damping constant, and this two are the extrema. Explictly we have
\begin{equation} k_{WD} = \frac{2\delta k \sqrt{k_1k_2}}{k_1 + k_2} \approx
\delta k, \label{kwd} \end{equation} for the weak decoherence mode, and
\begin{equation}
k_{SD} = k_1 + k_2 \approx 2k
\label{ksd}
\end{equation}
for the strong decoherence mode. One must note that while $k_{SD}$ is of the
same order as the individual damping constants $k_i$, the value of $k_{WD}$ can
be much lower. In the experiment with ions\cite{Science2} it was exactly this
lowering of the damping constant that was exhibited as an evidence of decoherence
``free'' subspaces. In the same experiment, one can see that the difference in
damping constants is much larger in the situation with an engineered noise
applied, since in this case the interaction with the environment is much closer
to the separability condition.

\section{Concluding Remarks}
\label{scremarks}
We have studied decoherence and dissipation free modes for systems of harmonic
oscillators. We discussed sufficient conditions for their existence. Although
theoretically simple, these conditions are very difficult to be implemented in
practical experiments. So we studied the slight deviations of this conditions,
and instead of decoherence free subspaces, we obtained {\emph{weak decoherence
modes}}. This suggests that weak decoherence subspaces can be used to store
quantum information for times much larger than individual carriers would be able
to, even without being rigorous DFS.

We compare DFS to the so called super and subradiance effects of a maser. In
fact, when in the last section we compare the damping constants for weak and
strong decoherence (eqs.~(\ref{kwd}) and (\ref{ksd})), this effect mimmimics
interference problem, where we are comparing the maximum and the minimum of a
certain quantity in which interference effects are recorded (in this case, the
damping constant).

It is important to stress that larger deviations of the rigorous conditions for
DFS preclude the existence of even weakly decoherence subspaces, by making the
decoherence time scales for such states smaller. However, one can conjecture
that this kind of mechanism is so general that whenever an experiment obtain
quantum mechanical results, it is testing some kind of DFS (\textit{e.g.} the
fullerenes experiment\cite{Arnetal}).

\begin{acknowledgments}
We gratefully acknowledge comments from A. N. Salgueiro.
This work was partly funded by FAPESP, CNPq and PRONEX (Brazil),
and Colciencias, DINAIN (Colombia). K.M.F.R. gratefully acknowledges
the Instituto de F\'\i sica, Universidade de S\~ao Paulo, for their
hospitality and PRONEX for partial support.
\end{acknowledgments}

\section*{Appendix: Realistic model of two oscillators in details}
The evolution superoperator for eq.(\ref{MasterdaSonia}) can be expressed as\cite{Sonia}
\begin{eqnarray}\label{rot}
\supop{U}\de{t}
& = & e^{\cjat\aponto\ad}e^{\cjbt\bp\bd}e^{\czt\bp\ad}
e^{\cczt\aponto\bd}
e^{\cqt{\op{a}_{1}}\bdp}
e^{\ccqt\pad{\op{a}_{2}}}e^{\cmbt\bd\bp}
e^{\ccmbt\pbd{\op{a}_{2}}}
\odot\nonumber\\
&&\odot  e^{\cmat\ad\aponto}
e^{\ccmat\pad{\op{a}_{1}}}
e^{\cqt\ad\bp}
e^{\ccqt\pad \op{a}_2}
\end{eqnarray}
where
\begin{eqnarray}
&&R=\frac{ k_2+
k_1}{2}+\frac{i\left(\omega_2+\omega_1\right)}{2},\quad c= k_2- k_1
+i\left(\omega_2-\omega_1\right),\quad
r=\sqrt{c^2+4k_3^{2}},\quad \Delta_{\pm}=
c\pm r \\
&&\cqt= 2k_3\left(1- e^{r\;t}\right)\left(
\Delta_{+}e^{r\;t}-\Delta_{-}\right)^{-1}\quad
{\textnormal{for}}\quad r\not=0 \\
&&e^{\cmat}=\frac{e^{-R\;t}}{2r}
e^{-\frac{r\;t}{2}}\left(\Delta_{+}e^{r\;t}-\Delta_{-}\right),
\quad e^{\cmbt}=e^{-2R\;t}e^{-\cmat} \\
&&\cjbt=\left(1+|\cqt|^{2}\right)
\left(\left|e^{\cmbt}\right|^{-2}\right)-1\\
&&\cjat=\left|e^{-\cmat}+\cqt^2 e^{-\cmbt}\right|^{2}
+\left(|\cqt|^2\right)\left(\left|
e^{\cmbt}\right|^{-2}\right)-1\\
&&\czt=-\cqt e^{-\left(\ccmat+\cmbt\right)}-\ccqt
\left( 1+|\cqt|^{2}\right)\left|e^{\cmbt}\right|^{-2}.
\end{eqnarray}
In this calculation neither the separability nor the degeneracy (even approximated) conditions
have been used so far.
For the sake of comparison we use the same initial condition of section \ref{s2osc} (eq.~(\ref{initialcondition})). In the general case
its time evolution is also given by
\begin{equation}
{\op{\rho}}(t)  = P(t) \ket{\psi(t)}\bra{\psi(t)}
+(1-P(t))\ket{0,0}\bra{0,0}
\end{equation}
but now the state $\ket{\psi(t)}$ is,
\begin{equation}
\ket{\psi(t)} =
\frac{(\cos(\theta)M_-(t)+\sin(\theta)e^{i\phi} Q(t)) \ket{1,0}
+(\sin(\theta)e^{i\phi} M_+(t)+\cos(\theta)Q(t))\ket{0,1}}{\sqrt{P(t)}},
\end{equation}
and its coefficient is given by
\begin{equation}
P(t) =
\left|
\cos(\theta)M_-(t)+\sin(\theta)e^{i\phi} Q(t)
\right|^2+
\left|
\sin(\theta)e^{i\phi} M_+(t)+\cos(\theta)Q(t)
\right|^2,
\end{equation}
where the functions $M_{\pm}(t)$ and $Q(t)$ are given by
\begin{equation}
M_{\pm}(t) = \frac{e^{-R t}}{2}
\left(
e^{- r t/2} (1\mp\frac{c}{r}) +e^{r t/2} (1\pm\frac{c}{r})
\right) , \quad
Q(t) = \frac{k_3}{r} e^{-R t}\left(e^{- r t/2}  - e^{r t/2}\right).
\end{equation}
Now, we assume slight deviations from degeneracy and separability,
that is, $\omega_1 = \omega-\delta \omega$, $\omega_2 = \omega+\delta \omega$, $k_3 =
\sqrt{k_1 k_2} -\delta k$, with $\delta \omega\ll \omega$, $\delta k\ll \sqrt{k_1 k_2}$.
Then, the state $\ket{\psi(t)}$ can be
approximated as
\begin{equation}
\ket{\psi(t)} = e^{-i\omega t}
\left(
\frac{\zeta_1\ket{1,0} +\xi_1\ket{0,1}}{\sqrt{P(t)}} e^{-(k_1+k_2)t}
      +\frac{\zeta_2 \ket{1,0}
      +\xi_2 \ket{0,1}}{\sqrt{P(t)}}
e^{-\frac{2\delta k \sqrt{k_1 k_2}}{k_1+k_2}t}
\right),
\label{genstate}
\end{equation}
where $\zeta_i,\xi_i$ do not have any temporal dependence and are
given by
\begin{eqnarray}
 \zeta_{\choice{1}{2}} & = &
\frac{(k_{\choice{1}{2}}\mp i\delta \omega)\cos(\alpha)
\pm\sqrt{k_1k_2}\sin(\alpha)e^{i\phi}}{k_1+k_2},
\\
\xi_{\choice{1}{2}} & = &
\frac{(k_{\choice{2}{1}}\pm i\delta \omega) e^{i\phi}\sin(\alpha)
        \pm\sqrt{k_1k_2}\cos(\alpha)}{k_1+k_2}.
\end{eqnarray}
In the general case it is not possible to find initial conditions
which are completely decoherence free. Nor it is possible to find two
orthogonal subspaces with very different characters in what
decoherence is concerned. However, we can choose the initial condition
as to have a minimal component either in a strong decoherence (SD) or in a
weak decoherence (WD) subspaces, by choosing, e.g.
\begin{equation}
tan(\alpha)_{\choice{\mbox{\tiny{SD}}}{\mbox{\tiny{WD}}}} =
\pm \sqrt{\frac{k_{\choice{2}{1}}}{k_{\choice{1}{2}}}},
\quad \phi_{\choice{\mbox{\tiny{SD}}}{\mbox{\tiny{WD}}}} = \pm \frac{\delta\omega}{k},
\end{equation}
where $k$ is some average dissipation constant. The corresponding
states, apart from a phase, can be written as
\begin{equation}\label{psiwd}
\ket{\psi_{\choice{{\mbox{\tiny{SD}}}}{\mbox{\tiny{WD}}}}} =
\frac{1}{\sqrt{k_1+k_2}}
\left(\sqrt{k_{\choice{1}{2}}}\ket{1,0}
\pm\sqrt{k_{\choice{2}{1}}}e^{\pm i \delta \omega/k}\ket{0,1} \right).
\end{equation}
If $\delta \omega\ll k_1, k_2$ the phase can be ignored. Moreover, if
$k_1=k_2$ then we have $k=k_1=k_2$ and the phase can be unaunambiguously
determined.
The weak decoherence wavefunction defines a mode which is robust
against decoherence. The damping constant of this mode can be read from eq.~(\ref{genstate}) as the value given in eq.~(\ref{kwd}). Analogously for the strong decoherence mode, with damping constant given by eq.~(\ref{ksd}).




\begin{thebibliography}{99}
\bibitem{comm} As must be expected, the experimental verification could
not find a completely isolated subspace as the simple theory foresee, but just a
sensible reduction on decoherence rate as compared to other subspaces. This can
also be viewed as a motivation to our work: to study the effects of small
deviations on the ideal situation.
\end{thebibliography}
\end{document}